\documentclass[aps,prd,twocolumn,groupedaddress,showpacs,nofootinbib]{revtex4}
\usepackage{graphicx}

\begin{document}

\title{Unified dark energy models\,: a phenomenological approach}

\author{V.F. Cardone}
\thanks{Corresponding author, email: {\tt winny@na.infn.it}}
\author{A. Troisi}
\author{S. Capozziello}
\affiliation{Dipartimento di Fisica ``E.R. Caianiello'', Universit\`a di Salerno and INFN, Sez. di Napoli, Gruppo Coll. di Salerno, via S. Allende, 84081 - Baronissi (Salerno), Italy}

\begin{abstract}

A phenomenological approach is proposed to the problem of universe accelerated expansion and of the dark energy nature. A general class of models is introduced whose energy density depends on the redshift $z$ in such a way that a smooth transition among the three main phases of the universe evolution (radiation era, matter domination, asymptotical de Sitter state) is naturally achieved. We use the estimated age of the universe, the Hubble diagram of Type Ia Supernovae and the angular size\,-\,redshift relation for compact and ultracompact radio structures to test whether the model is in agreement with astrophysical observation and to constrain its main parameters. Although phenomenologically motivated, the model may be straightforwardly interpreted as a two fluids scenario in which the quintessence is generated by a suitably chosen scalar field potential. On the other hand, the same model may also be read in the context of unified dark energy models or in the framework of modified Friedmann equation theories. 

\end{abstract}


\maketitle

\section{Introduction}

In the last few years, an increasing bulk of data has been accumulated leading to the emergence of a new cosmological scenario. The Hubble diagram of type Ia Supernovae (SNeIa) first indicated that the universe expansion is today accelerating \cite{Riess98,Perlm99}. The precise determination of first and second peaks in the anisotropy spectrum of cosmic microwave background radiation (CMBR) by the BOOMERanG and MAXIMA collaborations \cite{CMBR} strongly suggested that the geometry of the universe is spatially flat. When combined with the data on the matter density parameter $\Omega_M$, these results lead to the conclusion that the contribution $\Omega_X$ of dark energy is the dominant one, being $\Omega_M \simeq 0.3, \Omega_X \simeq 0.7$. This picture has been strenghtened by the recent determination of CMBR spectrum measured by the WMAP team \cite{WMAP}.

According to the standard recipe, pressureless cold dark matter and a homogenously distributed cosmic fluid with negative pressure, referred to as {\it dark energy}, fill the universe making up of order $95\%$ of its energy budget. What is the nature of this dark energy still remains an open and fascinating problem. The simplest explanation claims for the cosmological constant $\Lambda$ thus leading to the so called $\Lambda$CDM model \cite{Lambda}. Although being the best fit to most of the available astrophysical data \cite{WMAP}, the $\Lambda$CDM model is also plagued by many problems on different scales. If interpreted as vacuum energy, $\Lambda$ is up to 120 orders of magnitudes smaller than the predicted value. Furthermore, one should also solve the {\it coincidece problem}, i.e. the nearly equivalence of the matter and $\Lambda$ contribution to the total energy density.

As a response to these problems, much interest has been devoted to models with dynamical vacuum energy, dubbed {\it quintessence} \cite{QuintFirst}. These models typically involve scalar fields with a particular class of potentials, allowing the vacuum energy to become dominant only recently (see \cite{PR02,Pad02} for comprehensive reviews). Altough quintessence by a scalar field is the most studied candidate for dark energy, it generally does not avoid {\it ad hoc} fine tuning to solve the coincidence problem. On the other hand, a quintessential behaviour may also be recovered without the need of scalar fields, but simply by taking into account the effective contribution to cosmology of some (usually neglected aspects) of fundamental physics \cite{review}. A first tentative were undertaken showing that a universe with a non vanishing torsion field is consistent with SNeIa Hubble diagram and Sunyaev\,-\,Zel'dovich data on clusters of galaxies \cite{torsion}. The same quintessential framework can be obtained with the extension of Einstein gravity to higher order curvature invariants leading to a model which is in good agreement with the SNeIa Hubble diagram and the estimated age of the universe \cite{curv}. It is worth noting that these alternative schemes provide naturally a cosmological component with negative pressure whose origin is simply related to the geometry of the universe itself thus overcoming the problems linked to the physical significance of scalar fields.

Despite the broad interest in dark matter and dark energy, their physical properties are still poorly understood at a fundamental level and, indeed, it has never been shown that the two are in fact two different ingredients. This observation motivated the great interest recently devoted to a completely different approach to quintessence. Rather than the fine tuning of a scalar field potential, it is also possible to explain the acceleration of the universe by introducing a cosmic fluid with an exotic equation of state causing it to act like dark matter at high density and dark energy at low density. An attractive feature of these models is that they can explain both dark energy and dark matter with a single component (thus automatically solving the coincidence problem) and have therefore been referred to as {\it unified dark energy} (UDE) or {\it unified dark matter} (UDM).  Some interesting examples are the generalized Chaplygin gas \cite{Chaplygin}, the tachyonic field \cite{tachyons} and the condensate cosmology \cite{Bruce}.

It is worth noting that all the dark energy models (both with scalar fields or UDE) proposed up to now predict that the expansion of the universe is a two phase process\,: it is first determined by a matter\,-\,like term and it is then driven by the quintessence\,-\,like component towards an asymptotically de Sitter state. However, it is well established that there is also a third phase preceding these two, i.e. the radiation dominated era. It is thus interesting to look for a model which is able to predict a smooth transition from one phase to the following one in a natural way. 

A quite confused picture emerges from the previous discussion about dark energy and its nature. Many models have been proposed to succesfully reproduce the astrophysical observations available up to date, but they are so different each other that the mistery of the dark energy is far to be solved. Given the state of the art, a different approach to the problem is welcome. We think that a first step toward understanding the nature of dark side (dark energy and dark matter) of the universe is to explore phenomenological models which are able to reproduce what we observe. To this aim, we consider a single fluid whose energy density scales with the redshift in such a way that the radiation dominated era, the matter domination and the accelerating phase have been naturally achieved. We constrain the model parameters using the estimated age of the universe, the SNeIa Hubble diagram and the angular size\,-\,redshift relation of radio structures. Although phenomenologically motivated, nonetheless the model we propose may be physically interpreted in terms of an effective two fluids scenario with the dark energy component represented by a scalar field with a suitably chosen interaction potential. On the other hand, the models may also be considered as a new member of the UDE class. 

The plan of the paper is as follows. In Sect.\,II we present the main feature of the class of phenomenological models we propose. Some considerations on the allowed range for the model parameters lead us to restrict our attention (at least in this first analysis) to a particular class of models for the reasons we explain in Sect.\,III. The estimated age of the universe, the SNeIa Hubble diagram and the angular size\,-\,redshift relation for compact radio structures are used in Sects.\,IV, V and VI to test whether the model is a viable one and to constrain its parameters. Sect.\,VII is devoted to the interpretation of the model in terms of an effective two fluids model with a scalar field whose interaction potential is reconstructed. Finally, we summarize and conclude in Sect.\,VIII.

\section{A general class of models}

A phenomenological approach is proposed here to build a model which is able to fit the available data and leads to an accelerated expansion. To this aim, let us observe that most of the cosmological models (both with one or two fluids) predict that, during its evolution, the universe undergoes first a radiation dominated expansion (i.e. the energy density $\rho$ scales with the scale factor $R$ as $R^{-4}$), then a matter dominated phase (with $\rho \sim R^{-3}$) and finally a de\,Sitter\,-\,like expansion with the energy density asymptotically approaching a constant value. This consideration leads us to assume the following expression for the energy density\footnote{As yet said in the introduction, the model we are considering may be considered as composed by one or two fluids. Here, we prefer the single fluid interpretation, but the results we will obtain hold whatever is the number of fluids the model is made of. In Sect.\,VII, we will investigate the consequences of our results on the scalar field potential, if the model is considered as composed by two fluids.} which our model is made of\,:

\begin{equation}
\rho(R) = A \left ( 1 + \frac{s}{R} \right )^{\beta - \alpha} \ 
\left [ 1 + \left ( \frac{b}{R} \right )^{\alpha} \right ] 
\label{eq: rhor}
\end{equation}
with $0 < \alpha < \beta$, $s$ and $b$ (with $s < b$) two scaling factors and $A$ a normalization constant. For several applications, it is useful to rewrite the energy density as a function of the redshift $z$. Replacing $R = (1 + z)^{-1}$ in Eq.(\ref{eq: rhor}), we get\,:

\begin{equation}
\rho(z) = A \ \left ( 1 + \frac{1 + z}{1 + z_s} \right )^{\beta - \alpha} \ 
\left [ 1 + \left ( \frac{1 + z}{1 + z_b} \right )^{\alpha} \right ]
\label{eq: rhoz}
\end{equation}
having defined\,:

\begin{equation}
z_s = 1/s - 1 \ ,
\label{eq: defzs}
\end{equation} 
\begin{equation}
z_b = 1/b - 1 \ .
\label{eq: defzb}
\end{equation} 
It is quite easy to see that\,:

\begin{displaymath}
\rho \sim R^{-\beta} \ \ {\rm for} \ \ R << s \ ,
\end{displaymath}
\begin{displaymath}
\rho \sim R^{-\alpha} \ \ {\rm for} \ \ s << R << b \ , 
\end{displaymath}
\begin{displaymath}
\rho \sim const \ \ {\rm for} \ \ R >> b \ \ .
\end{displaymath}
Choosing $(\alpha, \beta) = (3, 4)$, the model we obtain is able to mimic a universe undergoing first a radiation dominated era (for $z >> z_s$), then a matter dominated phase (for $z_b << z << z_s$) and finally approaching a de\,Sitter phase with constant energy. This is just what we need. However, for sake of completeness, we discuss in this section the main properties of the model for the general case, i.e. with $(\alpha, \beta)$ not fixed\footnote{We may refer to this class of models as the {\it Hobbit models}. In Tolkien's trilogy {\it ``The Lord of the Rings''}, the Hobbits look like a ``mixture'' of the three main people of the book having the aspect of Men, (almost) the height of Dwarfs and pointed ears as Elfs. In the same way, our models behave as the three main fluids of the standard cosmological model.}.

As a preliminary step, let us recall the Friedmann equations \cite{CosmoBooks}\,:

\begin{equation}
H^2 + \frac{\kappa}{R^2} = \frac{8 \pi G}{3} \ \rho \ ,
\label{eq: fried1}
\end{equation}
\begin{equation}
2 \frac{\ddot{R}}{R} + H^2 + \frac{k}{R^2} = - 8 \pi G p \ ,
\label{eq: fried2}
\end{equation}
where $H = \dot{R}/R$ is the Hubble parameter, $p$ the pressure and the dot denoting the derivative with respect to $t$. From now on, we will consider only flat models so that $k = 0$. Inserting Eq.(\ref{eq: rhor}) into Eq.(\ref{eq: fried1}) for a flat universe and evaluating it today (where we set $R = 1$), we may express the normalization constant $A$ as\,: 

\begin{equation}
A = \frac{\rho_{crit}}{(1 + s)^{\beta - \alpha} \ (1 + b^{\alpha})}  
\label{eq: norm}
\end{equation}
with $H_0$ the present day Hubble constant and $\rho_{crit} = 3 H_0^2/8 \pi G$ the critical density. From now on, today evaluated quantities will be denoted by the label {\it ``o''}. Also the continuity equation\,:

\begin{equation}
\dot{\rho} + 3 H (\rho + p) = 0 \ .
\label{eq: cont}
\end{equation}
has to be taken into account.

Using the obvious relation $d\rho/dt = d\rho/dR \times dR/dt$ and the defintion of the Hubble parameter, it is immediate to get the following expression for the pressure\,:

\begin{equation}
p = - \frac{1}{3} \left ( R \frac{d\rho}{dR} + 3 \rho \right )
\label{eq: eqp}
\end{equation}
which holds whatever is the cosmological model\footnote{Actually, if the model contains more than a single fluid, this relation strictly holds only for the total pressure and the total energy density. However, it is still valid for each fluid if each one satisfies the continuity equation separately. This happens whenever the fluids are not interacting as in many quintessence models.} provided that $H$ is not vanishing everywhere (i.e. the universe is not stationary). Inserting Eq.(\ref{eq: rhor}) into Eq.(\ref{eq: eqp}), after some algebra we get\,:

\begin{equation}
w = \frac{[ (\alpha - 3) R + (\beta - 3) s] b^{\alpha} - [3 (R + s) + (\alpha - \beta) s ] R^{\alpha}}
{3 (R + s) (R^{\alpha} + b^{\alpha})} 
\label{eq: wr}
\end{equation}
with $w \equiv p/\rho$ as usual. The expression, as a function of $z$, may be easily obtained replacing $R = (1 + z)^{-1}$. The result for the case $(\alpha, \beta) = (3, 4)$ is\,:

\begin{equation}
w(z, \alpha = 3, \beta = 4) = \displaystyle{
\frac{\left [ \left ( \frac{1 + z}{1 + z_b} \right )^3 - 2 \right ] \frac{1 + z}{1 + z_s} - 3}
{3 \left ( 1 + \frac{1 + z}{1 + z_s} \right ) \left [ 1 + \left ( \frac{1 + z}{1 + z_s} \right )^3 \right ]}}
\ .
\label{eq: wzmodel}
\end{equation}  
Note that the barotropic factor $w$ strongly depends on the redshift $z$. In particular, Eq.(\ref{eq: wzmodel}) shows that\,:

\begin{figure}
\centering \resizebox{8.5cm}{!}{\includegraphics{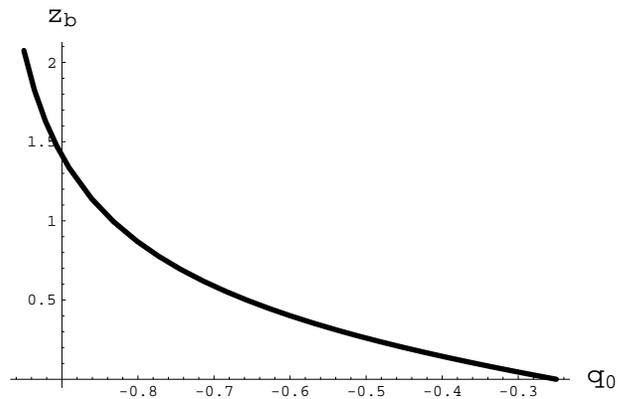}}
\caption{$z_b$ as function of $q_0$ for the model with $(\alpha, \beta) = (3, 4)$ and $z_s = 3454$.}
\label{fig: zb}
\end{figure}

\begin{displaymath}
w \sim 1/3 \ \ {\rm for} \ \ z >> z_s \ ,
\end{displaymath}
\begin{displaymath}
w \sim 0 \ \ {\rm for} \ \ z_b << z << z_s \ , 
\end{displaymath}
\begin{displaymath}
w \sim -1 \ \ {\rm for} \ \ z << z_b \ .
\end{displaymath}
Before discussing in more detail the behaviour of $w$ with $z$, let us first derive the expression of the deceleration parameter $q$. Combining the two Friedmann equations for a flat case, we easily get\,:

\begin{equation}
q(t) \equiv - \frac{\ddot{a} \ a}{\dot{a}^2} = \frac{1}{2} + \frac{3}{2} \frac{p}{\rho} \ .
\label{eq: defq}
\end{equation}
Inserting Eq.(\ref{eq: wr}) into this relation gives\,:

\begin{equation}
q = \frac{[(\alpha - 2) R + (\beta - 2) s] b^{\alpha} - [2 (R + s) + (\alpha - \beta) s] R^{\alpha}}
{2 (R + s) (R^{\alpha} + b^{\alpha})} \ ;
\label{eq: qr}
\end{equation}
inserting $R = 1$ gives the present day value as\,:

\begin{equation}
q_0 = \frac{(y - 1) \alpha + z_s [\alpha \ y - 2 (1 + y)] + (\beta - 4) (1 + y)}{2 (2 + z_s) (1 + y)}
\label{eq: qz}
\end{equation}
with $y = (1 + z_b)^{-\alpha}$. It is convenient to solve Eq.(\ref{eq: qz}) with respect to $z_b$ in order to express this one as a function of $q_0$ and $z_s$. It is\,:

\begin{equation}
z_b = \left [ \frac{\alpha (1 + z_s) + \beta - (2 + z_s) (2 q_0 + 2)}{\alpha - \beta + (2 + z_s) (2 q_0 + 2)} \right ]^{1/\alpha} - 1 \ .
\label{eq: solvezb}
\end{equation}
In Fig.\,\ref{fig: zb}, we plot $z_b(q_0)$ for the model with $(\alpha, \beta) = (3, 4)$ having fixed $z_s = 3454$ (see later for the motivation of this choice). For $q_0 \sim -0.5$, it is $z_b \sim 0.26$ so that such a model describes a universe which is dominated by a radiation\,-\,like fluid for $z >> 3454$, then its dynamical evolution is driven by a matter\,-\,like fluid until $z \sim 0.26$ when a term similar to the cosmological constant begins to dominate leading asymptotically to a de\,Sitter phase. This behaviour is quite similar to what is predicted by a cosmological model with dark matter and quintessence thus showing that the model we are considering is phenomenologically equivalent to the standard framework. Similar considerations refer to the models with other values of the parameters $(\alpha, \beta)$. 

Some straightforward physical considerations allow us to use Eq.(\ref{eq: solvezb}) to derive constraints on $q_0$. First, we note that $z_b$ is obviously a finite quantity so that we have to reject all values of $q_0$ which makes the right hand side of Eq.(\ref{eq: solvezb}) to diverge. Secondly, it is reasonable to assume that $z_b > 0$ since we need a matter dominated universe in the past so that structures can efficiently form. Imposing these two constraints, Eq.(\ref{eq: solvezb}) leads to the following condition\,:

\begin{equation}
q_{0,min} \le q_0 \le q_{0,max}
\label{eq: qzrange}
\end{equation}
with\,:

\begin{figure}
\centering \resizebox{8.5cm}{!}{\includegraphics{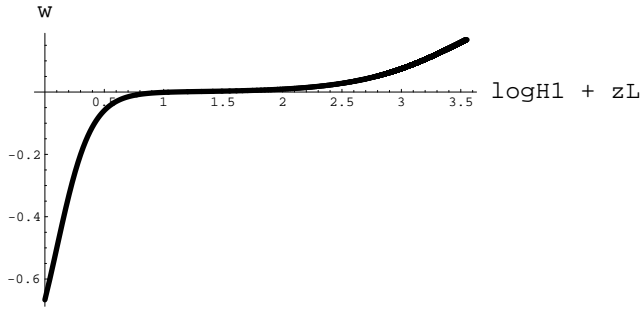}}
\caption{The barotropic factor $w$ as function of $log{(1 + z)}$ for the model with $(\alpha, \beta) = (3, 4)$ and $(q_0, z_s) = (-0.5, 3454)$.}
\label{fig: wz}
\end{figure}

\begin{equation}
q_{0,min} = \frac{1}{2} \left [ \frac{\beta - \alpha}{2 + z_s} - 2 \right ] \ ,
\label{eq: defqzmin}
\end{equation}
\begin{equation}
q_{0,max} = \frac{1}{2} \left [ \frac{\alpha z_s + 2 \beta}{2 (2 + z_s)} - 2 \right ] \ .
\label{eq: defqzmax}
\end{equation}
It is possible to see that both $q_{0,min}$ and $q_{0,max}$ are almost independent on $z_s$ for $z_s \in (1000, 4000)$ for the model with $(\alpha, \beta) = (3, 4)$ so that, with a very good approximation (more than $0.1\%$), we may fix\,:

\begin{displaymath}
q_{0,min}(\alpha = 3, \beta = 4) \simeq -1 \ , 
\end{displaymath}
\begin{displaymath}
q_{0,max}(\alpha = 3, \beta = 4) \simeq -0.25 \ .
\end{displaymath}

\begin{figure}
\centering \resizebox{8.5cm}{!}{\includegraphics{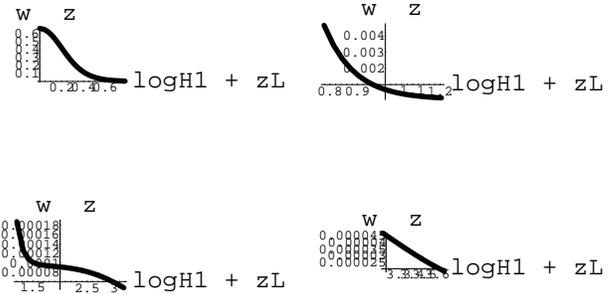}}
\caption{Derivative of the barotropic factor $w(z)$ vs $\log{(1 + z)}$ for the model with $(\alpha, \beta) = (3, 4)$ and $(q_0, z_s) = (-0.5, 3454)$. We consider four redshift ranges\,: $(0, 5)$ (top left), $(5, 15)$ (top right), $(15, 1500)$ (bottom left), $(1500, 4500)$ (bottom right). }
\label{fig: dwdz}
\end{figure}

Before constraining the model parameters with the available observations, we turn back again to the barotropic factor $w$ in order to qualitatively discuss some its interesting features. To this aim, we fix $(\alpha, \beta) = (3, 4)$ and $(q_0, z_s) = (-0.5, 3454)$ giving $z_b = 0.26$ and, in Fig.\,\ref{fig: wz}, we plot $w$ vs $log{(1 + z)}$. Note that $w(z)$ starts from a value near to that of a radiation dominated universe (i.e., $w = 1/3$), but it is exactly equal to that value only for $z \sim 100 z_s$ thus suggesting that the radiation dominated phase of the evolution takes place only at the very beginning. For $z < z_s$, $w$ is almost null (being $w = 0$ for $z \simeq 11$) coherently with the picture of a matter dominated universe. For $z < z_b$, $w$ quickly declines towards the asymptotic value $w = -1$ so that the pressure becomes negative and the universe enters a phase of accelerated expansion. For the chosen values of parameters $(\alpha, \beta)$ and $(q_0, z_s)$, it is $w(z = 0) = -2/3$. It is also interesting to look at the derivative of $w(z)$ that we plot in Fig.\,\ref{fig: dwdz}. The derivative is almost vanishing for most of the past history of the universe, being significantly different from zero only in the recent past because of the transition of the cosmic fluid from matter\,-\,like to quintessence\,-\,like regimes. This should suggest the use of an approximated equation of state as $p = w_{eff} \rho$ with $w_{eff}$ evaluated as a mean for $w(z)$ over the redshift. However, this procedure must be avoided since it could lead to serious systematic errors. Actually, looking at the relative variation $w^{-1} \times dw/dz$ shows that $w$ strongly depends on $z$ so that the introduction of a constant effective $w$ has no physical justification.

\subsection{Modified Friedmann equations ?}

Up to now, we have derived the main properties of the model by implicitely assuming that Eq.(\ref{eq: rhor}) describes the energy density of a single fluid accounting for both dark matter and dark energy. This phenomenologically motivated assumption may also be abandoned in favour of a different approach to the cosmic acceleration. As recently suggested, one might also think that, in the words of \cite{LSS03}, {\it the observed acceleration is not the manifestation of yet another new ingredient in the cosmic gas tank, but rather a signal of our first real lack of understanding of gravitational physics}. In this framework, one assumes that standard matter is the only component of a flat universe, while the Friedmann equation (\ref{eq: fried1}) is replaced by\,:

\begin{equation}
H^2 = H_0^2 g(x) \ ,
\label{eq: modfried}
\end{equation}
with $x = \rho_m/\rho_{crit}$ and $\rho_m$ scaling as usual (i.e., $\rho_m \propto R^{-3}$). The function $g(x)$ reduces to $x$ in the early stage of the universe evolution, while takes a different (nonlinear) form later. By suitably chosen $g(x)$, different models fitting the astrophysical data may be obtained, the most interesting ones being the (generalized) Cardassian model \cite{Cardassian} and the DGP gravity \cite{DGP}. 

Our phenomenological model may be interpreted in this framework provided that we choose\,:

\begin{eqnarray}
g(x) & = & \frac{(x/\Omega_{m,0})^{\beta/3}}{(1 + s)^{\beta - \alpha} (1 + b^{\alpha})} 
\left [ s + \left ( \frac{x}{\Omega_{m,0}} \right )^{-1/3} \right ] \times \nonumber \\
~ & ~ & \times \left [ b^{\alpha} + \left ( \frac{x}{\Omega_{m,0}} \right )^{-\alpha/3} \right ]   
\label{eq: ourg}
\end{eqnarray}
having used Eq.(\ref{eq: norm}) to fix the normalization constant $A$ and being $\Omega_{m,0} = \rho_{m,0}/\rho_{crit}$. In particular, for $(\alpha, \beta) = (3, 4)$, Eq.(\ref{eq: ourg}) reduces to\,:

\begin{equation}
g(x) = \frac{b^3 x}{\Omega_{m,0} (1 + s) (1 + b^3)} + g_{nl}(x) 
\label{eq: gspecial}
\end{equation}
with

\begin{equation}
g_{nl}(x) = \frac{1 + s (x/\Omega_{m,0})^{1/3} (1 + b^3 x/\Omega_{m,0})}{(1 + s) (1 + b^3)} \ .
\label{eq: gspecial2}
\end{equation}
Note that $g(x) \sim x$ in the early universe as expected\footnote{This is easy to show in the case $(\alpha, \beta) = (3, 4)$ remembering that $(x/\Omega_{m,0})^{1/3} = 1/R = 1 + z$.}. It is also interesting to compare Eqs.(\ref{eq: gspecial}) and (\ref{eq: gspecial2}) with the corresponding expression for a two fluids model composed of matter and cosmological constant having $g(x) \propto x + (1 - \Omega_{m,0})$. Also limiting to the case $(\alpha, \beta) = (3, 4)$, our model represents a generalization of the cosmological constant scenario since Eq.(\ref{eq: gspecial2}) reduces to this very special case only for $s = 0$ and adjusting $b$.  

In the following, we still prefer to interpret our model as a phenomenological one in the framework of a unified description of dark matter and dark energy. However, it is worth stressing that, since the astrophysical tests we will consider later are mainly sensitive to the shape of the Hubble function $H(z)$, the main results we will obtain are independent on what is the preferred physical meaning of the model among the different possibilities (UDE, matter plus scalar field or modified Friedmann equations).

\section{The space of parameters}

The general expression of energy density we are considering is characterized by five parameters which we may choose to be the two slopes $(\alpha, \beta)$, the scaling redshift $z_s$, the present day value of the deceleration parameter $q_0$ and the Hubble constant $H_0$ entering through the normalization coefficient $A$ in Eq.(\ref{eq: norm}). The astrophysical observations available up to date could be used to constrain the model parameters, but this is a daunting task given the large space of parameters to be search for. Actually, we have seen that models with $(\alpha, \beta) = (3, 4)$ are the most interesting ones since the energy density scales with $R$ as in the standard cosmological framework with matter and dark energy. Hence, hereinafter we will fix $(\alpha, \beta) = (3, 4)$. The space of parameters to search for is now significantly reduced since we have to consider only three of five quantities, namely $(z_s, q_0, H_0)$. Actually, $z_s$ marks the transiton of the fluid from radiation\,-\,like to matter\,-\,like regimes. In the standard framework, radiation and matter give the same contribution to the energy budget at $z_{eq}$ so that it is reasonable to assume that our fluid behaves as radiation much before this era. Henceforth, a possible choice could be $z_s \sim z_{eq}$. Fitting the $\Lambda$CDM model to the CMBR anisotropy spectrum, the WMAP collaboration has found (as best fit values) $z_{eq} = 3454$ \cite{WMAP} so that we fix $z_s$ to this value. However, we have checked that varying $z_s$ in the range $(1000, 5000)$ does not change the constraints on $(q_0, H_0)$ being all the tests we will discuss later fully degenerate with respect to $z_s$. This is not an unexpected result since the available observations probe a redshift range which is very far from $z_s$ whatever its exact value is. These considerations leave us with only two unknown quantities to constrain\,: the deceleration parameter $q_0$ and the Hubble constant $H_0$. In the following sections, we will use different astrophysical observations to check whether the model may be reconciled with them and to constrain these two parameters. 

\section{The age of the universe}

\begin{figure}
\centering \resizebox{8.5cm}{!}{\includegraphics{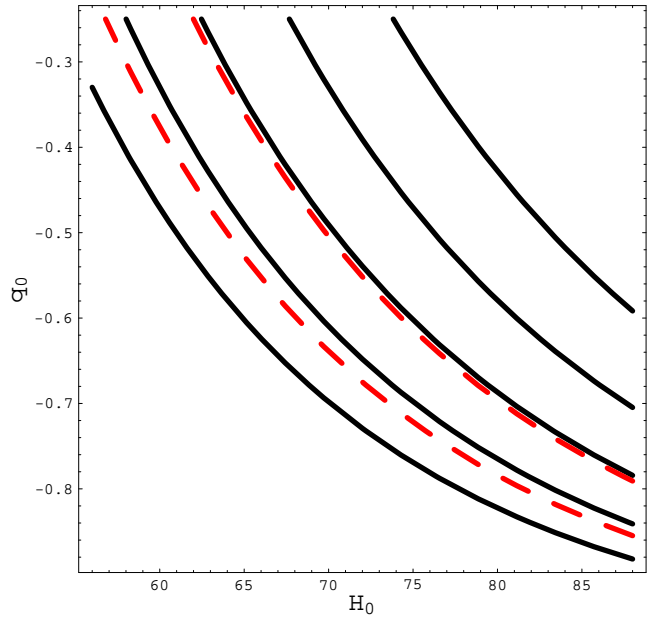}}
\caption{Age contours in the $(H_0, q_0)$ plane for the model with $(\alpha, \beta) = (3, 4)$ and $z_s = 3454$. The age values run from 11 (the upper line) to 15 Gyr (the lowest one) in steps of 1 Gyr. The dashed lines mark the region with $t_0 \in (13.1, 14.3) \ {\rm Gyr}$.}
\label{fig: age}
\end{figure}

In order to narrow the parameter space $(q_0, H_0)$, as a first step, we may compare the predicted age of the universe with the constraints coming from both astrophysical estimates and WMAP data. Inserting Eq.(\ref{eq: rhor}) into Eq.(\ref{eq: fried1}) with $A$ given by Eq.(\ref{eq: norm}), we get\,:

\begin{eqnarray}
t_0 & = & \sqrt{\frac{(1 + s)^{\beta - \alpha} \ (1 + b^{\alpha})}{H_0^2}} \times \nonumber \\ 
~ & ~ & \times \int_0^1{\frac{R^{-\beta} (R + s)^{\beta - \alpha} (R^{\alpha} + b^{\alpha})}{R} dR} \ .
\label{eq: time}
\end{eqnarray}
Eq.(\ref{eq: time}) is not analitically solvable, but may be easily integrated numerically provided that the values of parameters $(\alpha, \beta, z_s, q_0, H_0)$ have been given. Having yet fixed $(\alpha, \beta, z_s) = (3, 4, 3454)$, we have to choose only the ranges for the two remaining unknown quantities, $(q_0, H_0)$. We let $q_0$ vary in the full range determined before imposing the physically motivated constraint $z_b \ge 0$, i.e. we take $q_0 \in (-1, -0.25)$, while we examine models with $H_0 \in (56, 88) \ {\rm km \ s^{-1} \ Mpc^{-1}}$ since this is the $2 \sigma$ confidence range determined by the final result of the HST Key Project \cite{HSTKeyProj}. Eq.(\ref{eq: time}) is integrated over a grid in the $(q_0, H_0)$ plane with steps of $0.01$ in $q_0$ and $0.5$ in $H_0$ and interpolated for other values. In Fig.\,\ref{fig: age}, we plot the age contours in the $(H_0, q_0)$ plane. Superimposed, we show also the contours relative to the $3 \sigma$ confidence range determined by the WMAP data giving $t_0 \in (13.1, 14.3) \ {\rm Gyr}$, in good agreement with other independent astrophysical estimates. 

Fig.\,\ref{fig: age} shows that the age test is unable to put any constraint on the Hubble constant $H_0$, while put only a lower limit on $q_0$ imposing the constraint $q_0 \ge -0.85$. On the other hand, the test is a first evidence that the model we are considering is a reliable one since it predicts an age of the universe in agreement with the observational constraints.

\section{The SNeIa Hubble diagram}

\begin{figure}
\centering \resizebox{8.5cm}{!}{\includegraphics{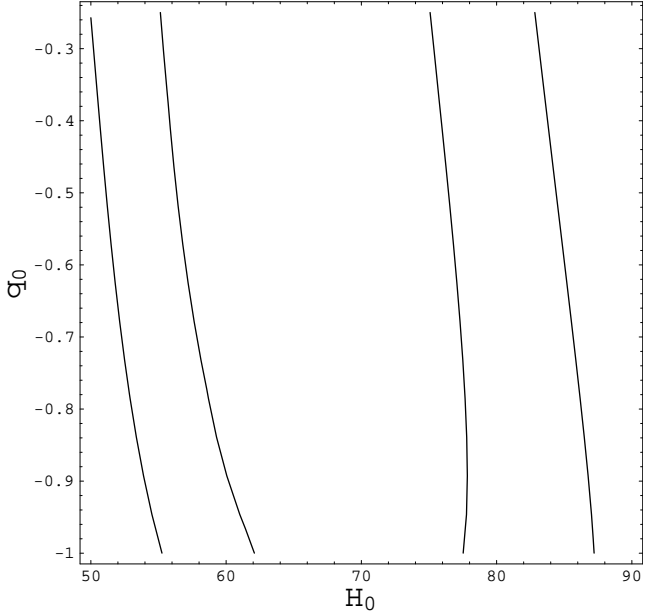}}
\caption{The 68 and 95$\%$ confidence ranges in the $(H_0, q_0)$ plane from fitting the model with $(\alpha, \beta, z_s) = (3, 4, 3454)$ to the SNeIa Hubble diagram.}
\label{fig: sneiafit}
\end{figure}

In order to further constrain the parameters $(q_0, H_0)$, we fit the model to the Hubble diagram of Type Ia Supernovae using the data recently released by the High\,-\,{\it z} Team \cite{HighZ} and the IfA Deep Survey \cite{IfA}. Following the method described in \cite{HighZ}, we minimize the quantity\,: 

\begin{eqnarray}
\chi^2 & = & \frac{1}{N - 2} \times \nonumber \\ 
~ & ~ & \times \sum_{i = 1}^{N}{\left [ \frac{\langle \log{H_0 d} \rangle_i - \log c d_L(z_i) + \log{h_{65}}}{\sigma_i} \right ]^2}
\label{eq: chidl}
\end{eqnarray}
with $\langle \log{H_0 d} \rangle_i$ the measured value of the distance (averaged over the different method used to compute it by the High\,-\,{\it z} Team), $d_L$ the dimensionless luminosity distance, $c$ the speed of light, $h_{65}$ the Hubble constant in units of $65 \ {\rm km \ s^{-1} \ Mpc^{-1}}$ and $\sigma_i$ the reported error and the sum is over the $N$ SNeIa observed. For our model with $(\alpha, \beta, z_s) = (3, 4, 3454)$, it is\,:

\begin{eqnarray}
d_L(z) & = & (1 + z) \ \sqrt{(1 + s) \ (1 + b^3)} \ \times \nonumber \\
~ & ~ & \times \int_{0}^{z}{d\zeta \left ( 1 + \frac{1 + \zeta}{1 + z_s} \right )^{-1/2} \ 
\left [ 1 + \left ( \frac{1 + \zeta}{1 + z_b} \right )^3 \right ]^{-1/2}} \ .
\label{eq: dl}
\end{eqnarray}
Note that $d_L$ depends on $q_0$ through $z_b$ since this is evaluated using Eq.(\ref{eq: solvezb}). We fit the model to the data using a sample of 162 SNeIa, 130 from Tonry et al. \cite{HighZ} and 23 from Barris et al. \cite{IfA}. These have been selected from a larger sample according to the two criteria $z > 0.01$ and $A_V < 0.5$ (being $A_V$ the absorption in the $V$ band) as in \cite{HighZ}. Note that we do not use the 42 SNeIa observed by the Supernova Cosmology Project (SCP) \cite{Perlm99} since their distance modulus have been estimated using a completely different approach with respect to those implemented by the High\,-\,{\it z} Team and also used by the IfA Deep Survey. Actually, Tonry et al. \cite{HighZ} have shown that inclusion of the SCP SNeIa does not alter the main results of fitting a model to the Hubble diagram so that we prefer to work with a homeogenous dataset even if this lowers the sample. We also take care of velocity uncertainties in estimating the supernova redshift adding 500 km/s divided by the redshift in quadrature to $\sigma_i$ following the prescription in \cite{HighZ}. The main results of the fitting procedure are presented in Fig.\,\ref{fig: sneiafit} where we plot the 68 and 95$\%$ confidence levels in the $(H_0, q_0)$ plane. 

Fig.\,\ref{fig: sneiafit} shows that fitting to the SNeIa Hubble diagram does not allow us to put constraints on $q_0$ since the contours are not closed along the $q_0$ direction. This is not true for $H_0$ so that we may get an estimate of this parameter. To this aim, since $\chi^2$ is of order 1, we first define the likelihood function as\,:

\begin{equation}
{\cal{L}}(q_0, H_0) \propto \exp{\{ -\chi^2/2 \}}
\label{eq: like}
\end{equation}
and then marginalize to get the two following functions\,:

\begin{equation}
{\cal{L}}_{q}(q_0) \propto \int{{\cal{L}}(q_0, H_0) \ dH_0} \ ,
\label{eq: likeqz}
\end{equation}

\begin{equation}
{\cal{L}}_{H}(H_0) \propto \int{{\cal{L}}(q_0, H_0) \ dq_0} \ .
\label{eq: likehubble}
\end{equation}
In Fig.\,\ref{fig: likeplot}, we plot the marginalized likelihoods normalized to their maximum values. From this plot, we see that the SNeIa test is completely degenerate with respect to $q_0$. The maximum is attained for $q_0 = -0.42$, but ${\cal{L}}_{q}$ varies less than $10\%$ over the full range $(-1, -0.25)$. On the other hand, ${\cal{L}}_{H}$ strongly depends on $H_0$ (with the maximum obtained for $H_0 = 64.3 \ {\rm km \ s^{-1} \ Mpc^{-1}}$) so that we get the following estimates for the Hubble constant\,:

\begin{displaymath}
H_0 \in (58.8, 72.3) \ {\rm km \ s^{-1} \ Mpc^{-1}}  \ \ (68\% CL) \ \ ,
\end{displaymath}
\begin{displaymath}
H_0 \in (53.1, 80.1) \ {\rm km \ s^{-1} \ Mpc^{-1}}  \ \ (95\% CL) \ \ . 
\end{displaymath}
This is in good agreement with the values obtained by the HST Key Project using various standard candles and from other independent techniques such as time delays in multiply imaged lens systems \cite{TimeDelay} and Sunyaev\,-\,Zeldovich clusters \cite{SZ}.

\section{The angular size\,-\,redshift test}

The relation between the (apparent) angular size and the redshift for compact radio structures in quasars and radio galaxies has been recently proposed as a possible cosmological test. Since radio data probe the redshift range bewteen  $z = 0.011$ and $z = 4.72$, it is clear that this test is potentially able to discriminate among different cosmological models breaking the degeneracy present in the lower redshift range proven by the SNeIa Hubble diagram. To see how this test works, let us consider an object having an intrinsic linear size $l$ and let $z$ be its redshift. The apparent angular size is (see, e.g., \cite{Gurv99})\,:

\begin{figure}
\centering \resizebox{8.5cm}{!}{\includegraphics{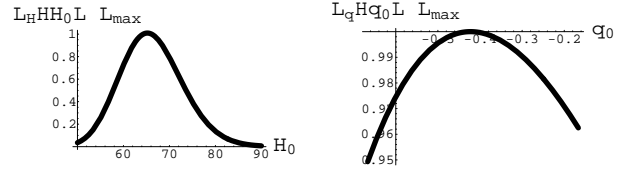}}
\caption{Marginalized likelihood functions (normalized to their maximum values) from fitting the model with $(\alpha, \beta, z_s) = (3, 4, 3454)$ to the SNeIa Hubble diagram.}
\label{fig: likeplot}
\end{figure}

\begin{eqnarray}
\theta(z) & = & \frac{l}{D_A(z)} = \frac{l (1 + z)^2}{D_L(z)} \nonumber \\
~ & = & \frac{l c}{H_0} \frac{(1 + z)^2}{d_L(z)} = \frac{D (1 + z)^2}{d_L(z)} 
\label{eq: thetaz}
\end{eqnarray}
with $D_A(z)$ the angular diameter distance, $D = l c/H_0$, the intrinsic angular size (in {\it mas}) and we have used the relation $D_L = (1 + z)^2 D_A$, with $D_L = c/H_0 \ d_L$ the luminosity distance yet introduced in Sect.\,5. It is worth stressing that Eq.(\ref{eq: thetaz}) implicitely assumes that the intrinsic linear size $l$ may be considered as a standard rod, i.e. it is the same whatever are the properties of the radio source. Actually, the validity of this hypothesis is still to be demonstrated, both observationally and on a theoeretical ground, and one should consider also a possible dependence of $l$ on the the total luminosity $L$ and/or on the redshift $z$. A simple way to parametrize these effects is to replace Eq.(\ref{eq: thetaz}) with the following phenomenological one \cite{Gurv99}\,:

\begin{figure}
\centering \resizebox{8.5cm}{!}{\includegraphics{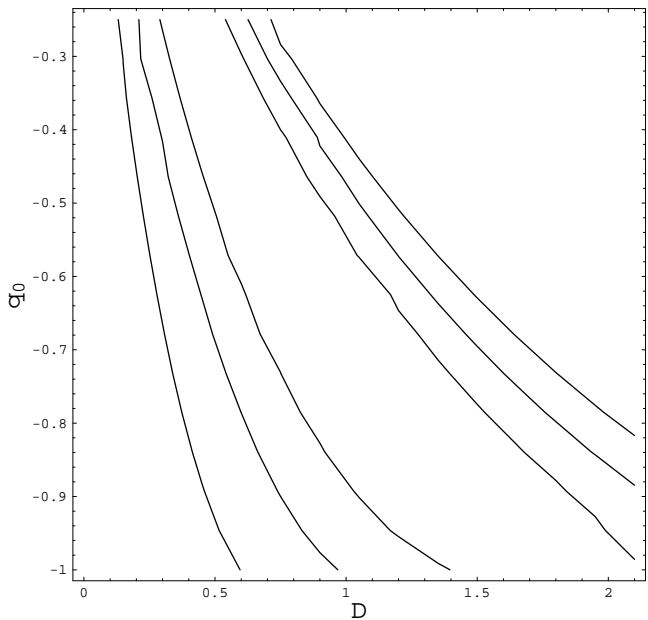}}
\caption{The 68, 95 and 99$\%$ confidence ranges in the $(D, q_0)$ plane from fitting the model with $(\alpha, \beta, z_s) = (3, 4, 3454)$ to the angular size\,-\,redshift relation for compact radio sources.}
\label{fig: angsizefit}
\end{figure}

\begin{equation}
\theta(z) = \kappa \ D \ L^{\gamma} \ (1 + z)^n \ D_A^{-1}(z) 
\label{eq: thetazbis}
\end{equation}
with $\kappa$ a normalization constant and $(\gamma, n)$ unknown parameters. Gurvits et al. \cite{Gurv99} have fitted this relation (assuming a $\Lambda = 0$ cosmological model) to the data coming from 145 radio sources (smoothed in 18 redshift bins) selected according to some selection criteria (see later) from a sample of 330 5 GHz VLBI sources. Their analysis shows that the estimates of the cosmological parameters obtained for different choices of $(\gamma, n)$ are consistent with each other within the errors. It is worth noting, however, that the uncertainties are quite large so that their result should be considered as an evidence of the degeneracy among different values of $(\gamma, n)$, not as a probe of the angular size\,-\,redshift test being independent on the choice of $(\gamma, n)$. Nonetheless, many authors \cite{AngSizeTest} usually assume $(\gamma, n) = (0, 0)$ so that we follow this approach and consider as unknown only the intrinsic angular size $D$ and the model parameters. In particular, being this test independent on the Hubble constant $H_0$ (since it has been included in the $D$ quantity) and having fixed as before $(\alpha, \beta, z_s)$, the deceleration parameter $q_0$ is the only model parameter we may constrain.

Following \cite{Gurv99}, we only select from the sample in Gurvits et al. the radio sources that have spectral index $\alpha_s \in (-0.38, 0.18)$ in order to reduce the intrinsic scatter in the angular size\,-\,redshift relation and smooth the data in (nearly) equally populated redshift bins. We miminize the quantity\,:

\begin{equation}
\chi^2 = \frac{1}{N - 2} \ \sum_{i}^{N}{\left [ \frac{\theta_{obs,i} - \theta_{mod}(z_i)}{\sigma_i} \right ]^2}
\label{eq: chiang}
\end{equation}
with $\theta_{obs,i}$ the observed value of the angular size in the redshift bin $z_i$, $\theta_{mod}(z_i)$ given by Eqs.(\ref{eq: dl}) and (\ref{eq: thetaz}) and $\sigma_i$ the uncertainty. Since the resulting $\chi^2$ for the best fit models is much lower than 1, we renormalize the errors in such a way that, for the best fit model, it is $\chi^2 = 1$. Although not statistically correct, this is the usual approach followed when dealing with data affected by likely overestimated errors. Moreover, this allows us to define marginalized likelihood functions proceeding in the same way as with Eqs.(\ref{eq: likeqz}) and (\ref{eq: likehubble}) in Sect.\,5. The main results of the angular size\,-\,redshift test are resumed in Figs.\,\ref{fig: angsizefit} and \ref{fig: likeplotang} showing, respectively, the $68\%$, $95\%$ and $99\%$ confidence ranges in the $(D, q_0)$ plane and the two marginalized likelihood functions (normalized to their maximum values). Because of the large errors affecting the data, we are still not able to constrain the value of the deceleration parameter. The best fit value turns out to be $q_0 = -0.88$, but the shape of the likelihood function only allows to put an upper limit, $q_0 \le -0.52$ at the $68\%$ confidence limit. On the other hand, the best fit value of the intrinsic angular size is $D = 0.81 \ mas$ and we get the following constraints\,:

\begin{figure}
\centering \resizebox{8.5cm}{!}{\includegraphics{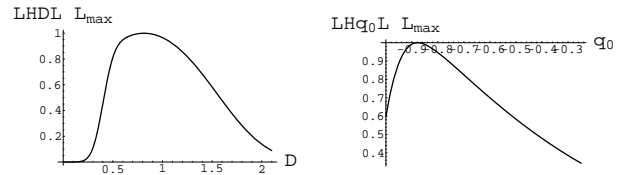}}
\caption{Marginalized likelihood functions (normalized to their maximum values) from fitting the model with $(\alpha, beta, z_s) = (3, 4, 3454)$ to the angular size\,-\,redshift relation of compact radio sources.}
\label{fig: likeplotang}
\end{figure}

\begin{displaymath}
D \in (0.44, 1.48) \ mas \ ,
\end{displaymath}
\begin{displaymath}
D \in (0.31, 2.00) \ mas \ .
\end{displaymath}
To investigate the effect of possible systematic errors and selection effects, we repeat the angular size\,-\,redshift test using a different sample comprising only ultracompact radio sources given by Jackson \cite{Jackson}. According to him, this dataset is more homogeneous than the one in \cite{Gurv99} and has also been corrected for any selection effect. Jackson also gives the values of $(z, \theta)$ to be used in the angular size\,-\,redshift test, while, following his prescription, the error on each of the six data points is estimated so that $\chi^2 = 1$ for the best fit model. We give the corresponding marginalized likelihood functions defined before in Fig.\,\ref{fig: likejack}. It is remarkable that the two likelihood functions are quite narrow so that it is possible to get constraints on both $D$ and $q_0$. The best fit value for the deceleration parameter is $q_0 = -0.64$, while the 68$\%$ and 95$\%$ turn out to be\,:

\begin{displaymath}
q_0 \in (-0.76, -0.54) \ , 
\end{displaymath}
\begin{displaymath}
q_0 \in (-0.83, -0.37) \ .
\end{displaymath}
These ranges do not contradict the results obtained before from the same test with different data. They are, however, significantly narrower. This is in line with what is found in ref.\,\cite{Jackson} for the $\Lambda$CDM model. It seems that the sample provided by Jackson allows to narrow the contraints on the fitting parameters (both for our model and the $\Lambda$CDM model) because of the lower dispersion of the data due to the removal of selection effects. 

\begin{figure}
\centering \resizebox{8.5cm}{!}{\includegraphics{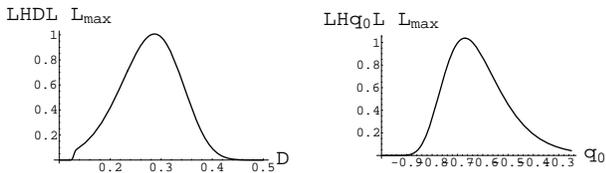}}
\caption{Same as Fig.\,\ref{fig: likeplotang} but using the data for ultracompact radio sources in \cite{Jackson}.}
\label{fig: likejack}
\end{figure}

Regarding the intrinsic angular size, we find $D = 0.28 \ mas$ as best fit value, while the corresponding 68$\%$ and 95$\%$ ranges are:\, 

\begin{displaymath}
D \in (0.22, 0.34) \ mas \ , 
\end{displaymath}
\begin{displaymath}
D \in (0.15, 0.39) \ mas \ .
\end{displaymath}
The intrinsic angular size turns out to be much smaller than what has been obtained using the sample in \cite{Gurv99}. However, this could be the result of the different selection criteria used to build the sample (ultracompact instead of compact sources). Furthermore, the apparent angular size is defined differently for the two samples (see the final remark in \cite{Jackson}).

\section{Reconstruction of the scalar field potential}

The model we have described and tested against some of the astrophysical observations available up to date has been proposed on a purely phenomenological basis. It is, however, interesting to observe that the same model has a straightforward interpretation in the standard framework of a universe made out of two fluids, namely the matter term and the (dominant) dark energy. To see this, let us consider again Eq.(\ref{eq: rhor}). We may rewrite it as $\rho = \rho_1 + \rho_2$ with\,:

\begin{equation}
\rho_1(R) = A \ b^{\alpha} \ R^{-\beta} \ (R + s)^{\beta - \alpha} \ ,
\label{eq: rhom}
\end{equation}
\begin{equation}
\rho_2(R) = A \ R^{\alpha - \beta} \ (R + s)^{\beta - \alpha} \ .
\label{eq: rho1}
\end{equation}
It is quite easy to see that\,:

\begin{displaymath}
\rho_1 \sim R^{-\beta} \ , \ \rho_2 \sim 0 \ \ {\rm for} \ R << s \ ,
\end{displaymath}
\begin{displaymath}
\rho_1 \sim R^{-\alpha} \ , \ \rho_2 \sim const \ \ {\rm for} \ R >> s \ ,
\end{displaymath}
\begin{displaymath}
\rho_1/\rho_2 = (b / R)^{\alpha} \sim 0 \ \ {\rm for} \ R >> b \ .
\end{displaymath}
For the model with $(\alpha, \beta) = (3, 4)$, $\rho_1$ scales with $R$ as a matter\,-\,like term\footnote{Here, as ``matter\,-\,like'' we mean here both radiation and matter.}, while $\rho_2$ as a quintessence\,-\,like fluid. Moreover, the matter\,-\,like term drives the evolution of the universe until $R < b$ (i.e., $z > z_b$), after which the quintessence\,-\,like term starts dominating. Motivated by this analogy, we will refer to $\rho_1$ as $\rho_m$ and to $\rho_2$ as $\rho_Q$. Let us now define the two dimensionless density parameters\,:

\begin{equation}
\Omega_m = \frac{\rho_m(R = 1)}{\rho_{crit}} = \frac{b^{\alpha}}{1 + b^{\alpha}} \ ,
\label{eq: om}
\end{equation}
\begin{equation}
\Omega_Q = \frac{\rho_Q(R = 1)}{\rho_{crit}} = \frac{1}{1 + b^{\alpha}} \ .
\label{eq: oq}
\end{equation}
$\Omega_m$ and $\Omega_Q$ may also be read as the density parameters of matter and quintessence, respectively. For $\alpha = 3$, this gives\,:

\begin{displaymath}
\Omega_m/\Omega_Q = b^3 \simeq 0.37 \rightarrow b \simeq 0.72 \rightarrow z_b \simeq 0.39 \ .
\end{displaymath}
On the other hand, using $q_0 = -0.64$ (as resulting from the angular size\,-\,redshift test using the Jackson data) and Eq.(\ref{eq: solvezb}), we get\,:

\begin{displaymath}
z_b \simeq 0.47 \rightarrow \Omega_m/\Omega_Q = (1 + z_b)^{-3} \simeq 0.32 
\end{displaymath}
in qualitative good agreement with the previous estimate. 

Let us now evaluate the barotropic factors of these two fluids. To this aim, we may use Eq.(\ref{eq: eqp}) to obtain\,:

\begin{equation}
w_m(R) = \frac{(\alpha - 3) R + (\beta - 3) s}{3 (R + s)} \ ,
\label{eq: wm}
\end{equation}
\begin{equation}
w_Q(R) = - \frac{3 (R + s) + (\alpha - \beta) s}{3 (R + s)} \ .
\label{eq: wq}
\end{equation}
Note that, to use Eq.(\ref{eq: eqp}), we have implicitly assumed that the two fluids do not interact so that the total pressure may be written as the sum of two single contributions. Actually, it is quite easy to verify that\,:

\begin{displaymath}
p_{tot} = p_m + p_Q = w_m \rho_m + w_Q \rho_Q 
\end{displaymath}
with $p_{tot} = w \rho$ and $w$ given by Eq.(\ref{eq: wr}). 

It is worth noting that, for $(\alpha, \beta) = (3, 4)$, Eqs.(\ref{eq: wm}) and (\ref{eq: wq}) reduces to\,:

\begin{displaymath}
w_m(R) = \frac{s}{3 (R + s)} \rightarrow w_m(R = 1) \simeq 0 \ ,
\end{displaymath} 	

\begin{displaymath}
w_Q(R) = -1 + \frac{s}{3 (R + s)} \rightarrow w_Q(R = 1) \simeq -1 \ ,
\end{displaymath} 
having used Eq.(\ref{eq: defzs}) and $z_s = 3454$. The present day values of the barotropic factors for the two fluids only depend on the $z_s$ and reduce to that of matter and cosmological constant, respectively, if $z_s \sim 10^{-3}$.

\begin{figure}
\centering \resizebox{8.5cm}{!}{\includegraphics{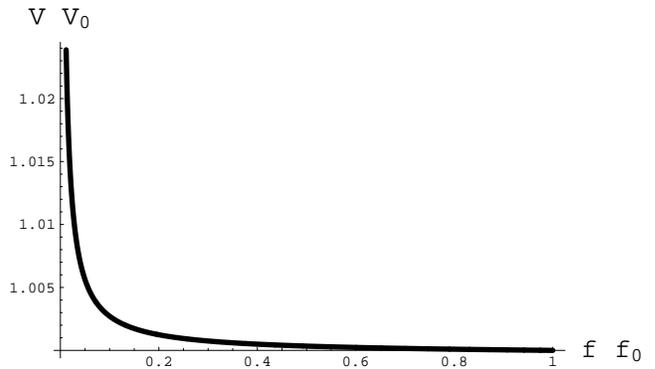}}
\caption{The scalar field potential for the model with $(\alpha, \beta) = (3, 4)$ and $(z_s, q_0) = (3454, -0.64)$. Here, $\phi_0$ and $V_0$ are the present day values of the scalar field and of the potential.}
\label{fig: vphi}
\end{figure}

Let us now concentrate on the quintessence\,-\,like term. In the standard framework, the quintessence fluid is generated by a scalar field $\phi$ rolling down its potential $V(\phi)$. These quantities are then related to the energy density $\rho_Q$ and the barotropic factor $w_Q$ as follows \cite{Pad02}\,:

\begin{equation}
\left \{
\begin{array}{ll}
\rho_Q = \displaystyle{\frac{1}{2} \dot{\phi}^2 + V(\phi)} \\
~ \\
w_Q = \displaystyle{\frac{1 - 2 V/\dot{\phi}^2}{1 + 2 V/\dot{\phi}^2}} \\ 
\end{array}
\right .
\label{eq: sys}
\end{equation}
which, using Eq.(\ref{eq: fried1}), can be solved\footnote{The reconstruction of the scalar field potential from the data is always possible provided that the right set of equations are used and the data are of sufficient quality to prevent strong degeneracies in the potential reconstruction. An interesting example may be inferred from the equations presented in \cite{Wett} where a model with varying scales and couplings is discussed.} with respect to $\phi$ and $V$ to give\,:

\begin{equation}
\left \{
\begin{array}{ll}
\phi = \displaystyle{\sqrt{\frac{3}{4 \pi G}} \ \int_{0}^{R}
{\left [ \frac{R'^{\alpha}}{R'^{\alpha} + b^{\alpha}} \left ( 1 + \frac{1 - w_Q}{1 + w_Q} \right ) 
\right ]^{-1} \frac{dR'}{R'}}} \ , \\
~ \\
V = \displaystyle{\left ( 1 + \frac{1 - w_Q}{1 + w_Q} \right )^{-1} \ \frac{1 - w_Q}{1 + w_Q} \ \rho_Q \ .}
\end{array}
\right .
\label{eq: invsys}
\end{equation}
Solving numerically this system allows to get the scalar field potential $V(\phi)$ that we plot in Fig.\,\ref{fig: vphi} for the model with $(\alpha, \beta, z_s) = (3, 4, 3454)$. In this plot, we have fixed $q_0 = -0.64$ as suggested by the best fit to the angular size\,-\,redshif test with the Jackson data. However, we have found that the shape of $V/V_0$ vs $\phi/\phi_0$ does not depend on $q_0$, while the deceleration parameter determines the present day values of both the scalar field ($\phi_0$) and the potential ($V_0$).  

The most striking result is that, although a monotonically decreasing function of the scalar field as expected, the variation of the potential with respect to $V_0$ is quite small, less than $2\%$ over the full redshift range. In other words, the reconstructed scalar field potential is almost the same as that of the standard cosmological constant. This result could also be expected from Eq.(\ref{eq: wq}). For the model with $(\alpha, \beta) = (3, 4)$, the departures of $w_Q$ from the the cosmological constant value $w_{\Lambda} = -1$ are driven by the term $(1 + R/s)^{-1}$ which is only slowly varying since $R >> s$ over almost the full redshift range. It is worth stressing, however, that $w_Q$ is never exactly equal to the cosmological constant value. This can also be seen considering its derivative with respect to the redshift being\,:

\begin{equation}
\frac{dw_Q}{dz} = \frac{(\beta - \alpha) (1 + z_s)}{(2 + z + z_s)^2} \ .
\label{eq: dwqdz}
\end{equation}
For $(\alpha, \beta) = (3, 4)$ and whatever is the value of $z_s$, this quantity does not identically vanish so that, even if the scalar field potential is slowly varying, the corresponding energy density may not be considered that of a cosmological constant (for which $V(\phi)$ does not depend at all on $\phi$). Therefore, some of the problems connected with the $\Lambda$ phenomenology are not present for our parametrization. On the other hand, a combined analysis of most of the available data gives the standard $\Lambda$CDM model as best fit \cite{WMAP} so that it is not surprising that using a similar dataset individuates as best fit model among our class the one that best matches the $\Lambda$CDM one.  

\section{Conclusions}

The concordance cosmological model assumes that pressureless cold dark matter and the dominant quintessence field drive the evolution of the universe leading to the observed accelerated expansion. As an alternative to this picture, unified dark energy models have been proposed where a single cosmic fluid acts both as dark matter and dark energy depending on the value of the energy density. 

Here we have presented a general class of models and tested it against the astrophysical observations available up to date. The starting point is the assumption that the fluid energy density $\rho$ depends on the scale factor $R$ as shown in Eq.(\ref{eq: rhor}). The model is characterized by five parameters, but we have limited our attention to those models where the energy density smoothly interpolates among the three main phases of the universe evolution, i.e. a radiation dominated era followed by matter domination and an asymptotic de Sitter state. This ansatz is motivated by the phenomenology we observe since every consistent picture of the universe evolution predicts the onset of the three different phases we have quoted above. The model has been tested against some astrophysical observations, namely the age of the universe, the SNeIa Hubble diagram and the angular size\,-\,redshift relation for compact radio structures. The succesful results of these tests is a strong evidence of the reliability of this phenomenological approach. It is worth noting that our results may be interpreted both in the frame of UDE models and in the standard picture of a two fluid universe made of matter and dark energy. In this latter case, if a scalar field $\phi$ is assumed to be the origin of this component, the interaction potential $V(\phi)$ may be directly reconstructed from the observations without any a priori hypothesis on its form. 

Although the model has been shown to successfully fit the available observations, further analysis is still needed. 

First, there are other tests that can be used to better constrain the model parameters. In particular, we have not discussed the growth of perturbations which determines the CMBR anisotropy spectrum. Since the energy density $\rho$ scales as the matter term in $\Lambda$CDM model during the structure formation era, perturbations should grow in the same manner so that the fit to the observed CMBR spectrum is likely to be successful. However, this naive expectation has to be carefully checked since it depends on how we interpret the model. Actually, this qualitative picture should indeed be valid if Eq.(\ref{eq: rhor}) is meant as a phenomenological description of a two fluids scenario since, in this case, the quintessence\,-\,like fluid almost vanishes during the epoch of structure formation and therefore everyting works as in the usual scenario. On the other hand, if the UDE interpretation is preferred, one has to explicitely solved the perturbation equations by using explicitely Eq.(\ref{eq: rhor}) for the matter enery density. We are, however, confident that the main results are not changed for the model with $(\alpha, \beta) = (3, 4)$ since, in this case, Eq.(\ref{eq: rhor}) reduces to the usual expression $(\rho \propto R^{-3})$ in the structure formation epoch. Finally, a different approach has to be considered if the interpretation of the model in the framework of modified Friedmann equations. In this case, one should first obtain the corresponding modified Newtonian potential and then work out the consequences on the equations describing the growth of perturbation to finally obtain a coherent description of the structure formation process and of its imprint on the CMBR anisotropy spectrum. 

Secondly, in this analysis, we have held fixed the two slopes parameters $(\alpha, \beta)$ to the values they must have to perfectly mimic the scaling of the energy density with the scale factor $R$ during the radiation and matter dominated era. However, it is still possible that other values of $(\alpha, \beta)$ fit the astrophysical data we have considered. It is thus interesting to repeat the same analysis performed here also varying these parameters. To this aim, however, more data are welcome since adding more parameters may introduce strong degeneracies among some of them. In order to solve these problems, one should also consider the observations of large scale distribution of matter as measured by the two major ongoing galaxy surveys (the 2dFGRS \cite{2dFGRS} and the SDSS \cite{SDSS}) or the data on the Lyman $\alpha$ forest \cite{Lyalpha}.

As a final remark, we want to stress again the underlying philosophy of this work. Contrary to the usual approach of proposing a theory and then test it against observations, we have preferred to start from the phenomenology we observe to investigate what are the main features a theory should have to give a consistent and realistic picture of the universe. 

\acknowledgments{We warmly thank Leonid Gurvits for help with his compilation of radio data. We also acknowledge Sante Carloni and Mauro Sereno for the interesting discussions on the topic. Finally, we want to thank an anonymous referee for his comments that have helped us to improve the paper.}

\end{document}